\documentclass[conference]{IEEEtran}

\ifCLASSINFOpdf
  \usepackage[pdftex]{graphicx}
\else
\fi

\usepackage{amssymb}
\usepackage{amsmath}
\usepackage{balance}
\usepackage{subcaption}
\usepackage{gensymb}
\usepackage{cite}
\usepackage{booktabs}
\begin{document}
%
\title{Emitter Identification Using CNN IQ Imbalance Estimators}

\author{\IEEEauthorblockN{Lauren J. Wong, William C. Headley, Alan J. Michaels}
\IEEEauthorblockA{Hume Center for National Security and Technology \\
Virginia Polytechnic Institute and State University\\
Email: \{ ljwong, cheadley, ajm \} @vt.edu }}

\maketitle

\begin{abstract}
Specific Emitter Identification is the association of a received signal to a unique emitter, and is made possible by the naturally occurring and unintentional characteristics an emitter imparts onto each transmission, known as its radio frequency fingerprint. 
This work presents an approach for identifying emitters using Convolutional Neural Networks to estimate the IQ imbalance parameters of each emitter, using only raw IQ data as input. 
Because an emitter's IQ imbalance parameters will not change as it changes modulation schemes, the proposed approach has the ability to track emitters, even as they change modulation scheme. 
The performance of the developed approach is evaluated using simulated quadrature amplitude modulation and phase-shift keying signals, and the impact of signal-to-noise ratio, imbalance value, and modulation scheme are considered. 
Further, the developed approach is shown to outperform a comparable feature-based approach, while making fewer assumptions and using less data. 
\end{abstract}

%
\IEEEpeerreviewmaketitle

\section{Introduction}
\label{intro}
Specific Emitter Identification (SEI) is the act of assigning an emitter with a received signal, using a database of radio frequency (RF) features. 
SEI algorithms are often used in military settings for emitter tracking \cite{talbot2003specific}, and have also become a powerful tool for use in cognitive radio applications to enforce Dynamic Spectrum Access (DSA) rules \cite{Park2014}.

Current state-of-the-art SEI systems rely on the measurement of pre-determined and expert-defined signal features, which are then clustered by emitter for identification \cite{talbot2003specific}. 
However, the extraction of expert features often requires considerable pre-processing of the raw signal data, such as synchronization, carrier tracking, demodulation, and signal-to-noise ratio (SNR) estimation, in addition to the computational cost of measuring or estimating the expert features. 
Further, these pre-determined features are often only accurate over a narrow range of parameters and require accurate and consistent measurement or estimation, in order to ensure quality SEI performance.

Features often used in traditional SEI algorithms are either taken from the transient portion or the steady-state portions of the received signal. 
When using features extracted from the transient signal, SEI performance relies heavily on the accuracy and consistency of the transient detection and removal process, as this directly affects the quality of the features \cite{7395025}.
Though using features extracted from the steady-state portion of the received signal is generally more practical, expert features used to describe the steady-state signal often have their own limitations. 
For example, wavelet-based techniques are heavily impacted by the choice of wavelet function \cite{6104141}. 
Preamble-based techniques fail in the case where the received signal does not have a pre-defined preamble \cite{7395025}. 
Further, techniques analyzing the cyclostationary features of a signal are often inconsistent in the presence of frequency or phase uncertainties \cite{kim_2008}.

While neural networks have been used for emitter identification to perform the classification stage, taking in pre-defined features as input \cite{8008202}, this work investigates the ability to perform emitter identification using neural networks to perform the feature extraction/estimation stage. 
In this work, an approach using Convolutional Neural Networks (CNNs) to extract an expert feature, transmitter IQ imbalance, is developed and analyzed. 
Further, using the developed CNN IQ imbalance estimators, an approach is presented to identify emitters across numerous modulation schemes, assuming the modulation class of the signal is known or the ability to determine the modulation class. 
Under no circumstances is it assumed that the CNN input is adjusted for synchronization, carrier-tracking, SNR estimation, or via demodulation.

Simulation results show that CNNs can be used to estimate both gain and phase imbalance. 
However, the CNN estimates of gain imbalance have less bias and variance than the CNN estimates of phase imbalance, indicating that more input samples or a more sophisticated network is needed to estimate phase imbalance than gain imbalance. 
Further, performance analysis of the developed SEI approach shows the ability to identify emitters by their gain imbalance only, with higher accuracy than a comparable feature-based approach. 

This paper is organized as follows: 
Section \ref{sec:est_iqi} discusses transmitter IQ imbalance and develops an appropriate signal model to be used in the generation of the simulated data used for training and testing of the approach. 
In Section \ref{sec:est}, the models designed to estimate IQ imbalance are described and thoroughly analyzed, using quadrature amplitude modulation (QAM) and phase-shift keying (PSK) test signals. 
Next, Section \ref{sec:estSEI} presents the SEI approach using the developed CNN IQ imbalance estimators and shows simulation results, including comparison to an existing feature-based approach. 
Finally, Section \ref{sec:fw} concludes the work.

\section{Transmitter IQ Imbalance}\label{sec:est_iqi}
\subsection{Causes and Implications}\label{sec:est_iqiCause}
Transmitter-induced frequency-independent IQ imbalance is caused by non-idealities in the local oscillators, mixers, and differential pair wiring of the transmitter which cause the in-phase and quadrature components of the modulator to be non-orthogonal. 
The result is the real and imaginary components of the complex signal interfering with each other. 
In addition to potentially degrading the performance of the transmitter, IQ imbalance can also be used as an identifying feature when performing SEI techniques.

\begin{figure}[t]
\vspace{-10pt}
  \centering
 \includegraphics[width=1.0\linewidth]{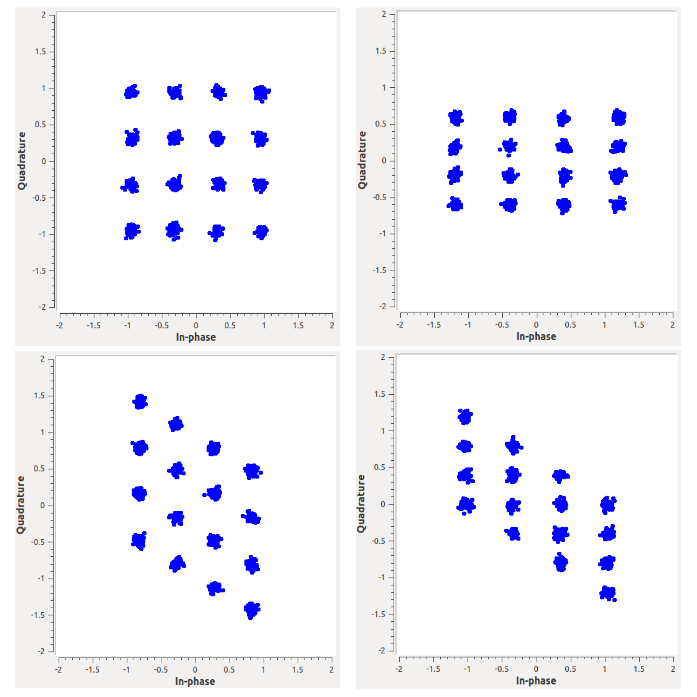}
	\caption{The result of transmitter IQ imbalance applied to the in-phase component of a 16QAM signal in the constellation diagram, SNR = 20dB. Top Left: no imbalances. Top Right: phase imbalance = $30^\circ$, gain imbalance = 0. Bottom Left: phase imbalance = 0, gain imbalance = 0.9. Bottom Right: phase imbalance = $30^\circ$, gain imbalance = 0.9.}
	\label{fig:imbalance_constellation}
    \vspace{-10pt}
\end{figure}

IQ imbalance in the constellation diagram, shown with exaggerated imbalance values and after demodulation for clarity, is shown for 16QAM in Figure \ref{fig:imbalance_constellation}. 
The result of a phase imbalance on a signal, shown in the lower left constellation, is a rotation of the real component of the symbols in the IQ plane. 
The result of a gain imbalance on a signal, shown in the upper right constellation, is a stretching or contracting of the of the real component of symbols along the in-phase axis.  
However, in many systems, it may be impractical to obtain the symbols, such as in a blind system where synchronization cannot be assumed. 
Given this, the proposed approach uses raw IQ as input, eliminating the need for demodulation, used in many traditional methods \cite{4389078}.  

\begin{figure}[t]
\vspace{-10pt}
  \centering
  \includegraphics[width=1.0\linewidth]{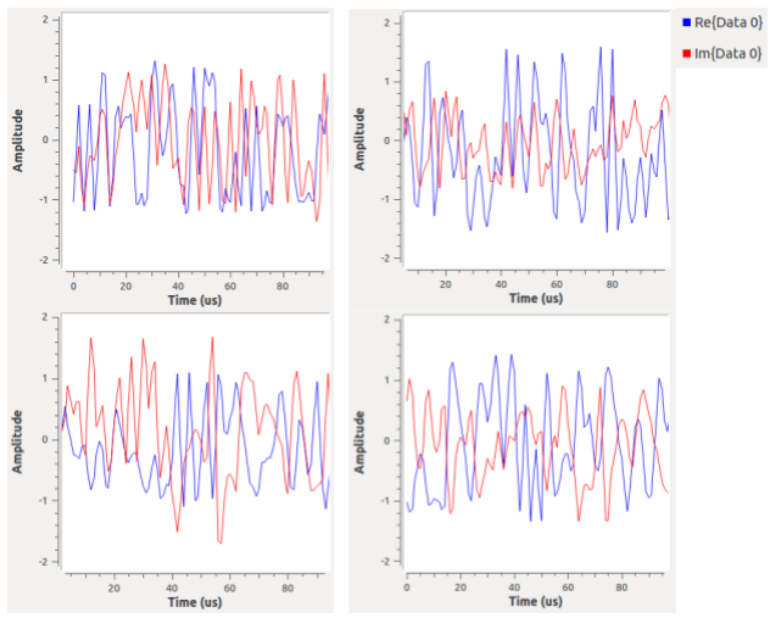}
	\caption{The result of transmitter IQ imbalance applied to the in-phase component of a 16QAM signal in the time domain, SNR = 20dB. Top Left: no imbalances. Top Right: phase imbalance = $30^\circ$, gain imbalance = 0. Bottom Left: phase imbalance = 0, gain imbalance = 0.9. Bottom Right: phase imbalance = $30^\circ$, gain imbalance = 0.9.}
    \label{fig:imbalance_time}
    \vspace{-10pt}
\end{figure}

IQ imbalance in the time domain is shown for 16QAM in Figure \ref{fig:imbalance_time}. 
The result of a gain imbalance on a signal in the time domain is an increase or decrease in the amplitude of the real component of the signal. 
The result of a phase imbalance on a signal in the time domain is a shifting of the phasor of the real component of the signal. 
To the human eye, a phase imbalance is much harder to see than a gain imbalance, though both become hard to detect at low SNR. 
However, as will be shown in Section \ref{sec:est_results}, using the  learned features, CNNs are able to identify small differences between sets of samples to estimate these imbalances, given enough samples and reasonable SNR values.


\subsection{Signal Model}\label{sec:est_iqiModel}
This work assumes only frequency-independent IQ imbalance. 
Though most modern communications systems are affected by frequency dependent IQ imbalance, frequency independence is often assumed in the existing literature, for simplicity \cite{Li2014}. 
Frequency independent IQ imbalance is also a valid approximation for imbalanced narrowband systems and imbalance due to the analog components of emitters \cite{inti_2017,Li2014}. 
Without loss of generality, all imbalances are modeled on the in-phase component of the modulated signal before transmission through an additive white Gaussian noise (AWGN) channel \cite{HsuSheen}, as follows: 

Consider the baseband signal 
\begin{equation}
x(t) = x_i(t) + jx_q(t),
\end{equation}
where $x_i(t)$ and $x_q(t)$ are real-valued time-varying baseband signals. 
An IQ modulator with imbalance modulates this baseband signal to its bandpass equivalent through
\begin{equation}
x(t) = (1 + \alpha)\cos(2\pi f_0 t + \theta) x_i(t) - j\sin(2 \pi f_0 t) x_q(t), 
\end{equation}
where $f_0$ is the carrier frequency, the transmitter's gain imbalance is represented by $\alpha$, and the transmitter's phase imbalance is represented by $\theta$, such that the ideal transmitter, with no IQ imbalance, has $\alpha = 0$ and $\theta = 0^\circ$. 
Transmission through an AWGN channel gives the received signal
\begin{align}
y(t) = \mathbb{R} \bigg\{ \sum_{k = -\infty}^\infty (1 &+ \alpha)\cos(2\pi f_0 t + \theta) x_{k_i}(t) \\
&- j\sin(2 \pi f_0 t) x_{k_q}(t) \bigg\} + n(t) \nonumber
\end{align}
where $n(t)$ is a zero mean white Gaussian noise process \cite{Proakis,713223}.

\subsection{Dataset Generation}\label{sec:est_data}
Using the signal model described above, all data used in the following simulations was generated with the open-source \emph{gr-signal\_exciter} module in GNURadio \cite{clark_gnu}. 
Though gain and phase imbalance values for real systems are not easily found, prior works in IQ imbalance estimation and compensation use test values ranging from 0.02 to 0.82 for absolute gain imbalance and from $2^\circ$ to $11.42^\circ$ for phase imbalance, with most works used test values on the orders of 0.05 and $5^\circ$ for gain and phase imbalance respectively \cite{4389078, HsuSheen, 4102451, 5153348, 293640, 4753809, 713223}. 
Therefore, QAM signals of orders 8, 16, 32, and 64 and PSK signals of orders 2, 4, 8, and 16 were simulated with  linear gain imbalances between [-0.9, 0.9], uniformly distributed, and phase imbalances between $[-10^\circ, 10^\circ]$, uniformly distributed, in order to incorporate all possible offset values one might see in a real system.
Additionally, frequency offsets between [-0.1, 0.1] times the sample rate, uniformly distributed, were simulated and the simulated signal was sampled between [1.2, 4] times Nyquist, uniformly distributed, in order to simulate the effects of an imperfect signal detection stage \cite{steve_milcom}.
The sampled signal was passed through a root-raised cosine filter with a roll-off factor of 0.35 and the signal normalized so that the average symbol power is 1dB. 
Finally, white Gaussian noise was added such that all signals had SNRs between [0dB, 25dB], uniformly distributed. 

\section{CNN IQ Imbalance Estimators}\label{sec:est}
\subsection{Model Design, Training, and Evaluation}\label{sec:est_model}
\begin{figure}[t]
	\centering
	\includegraphics[width=1.0\linewidth]{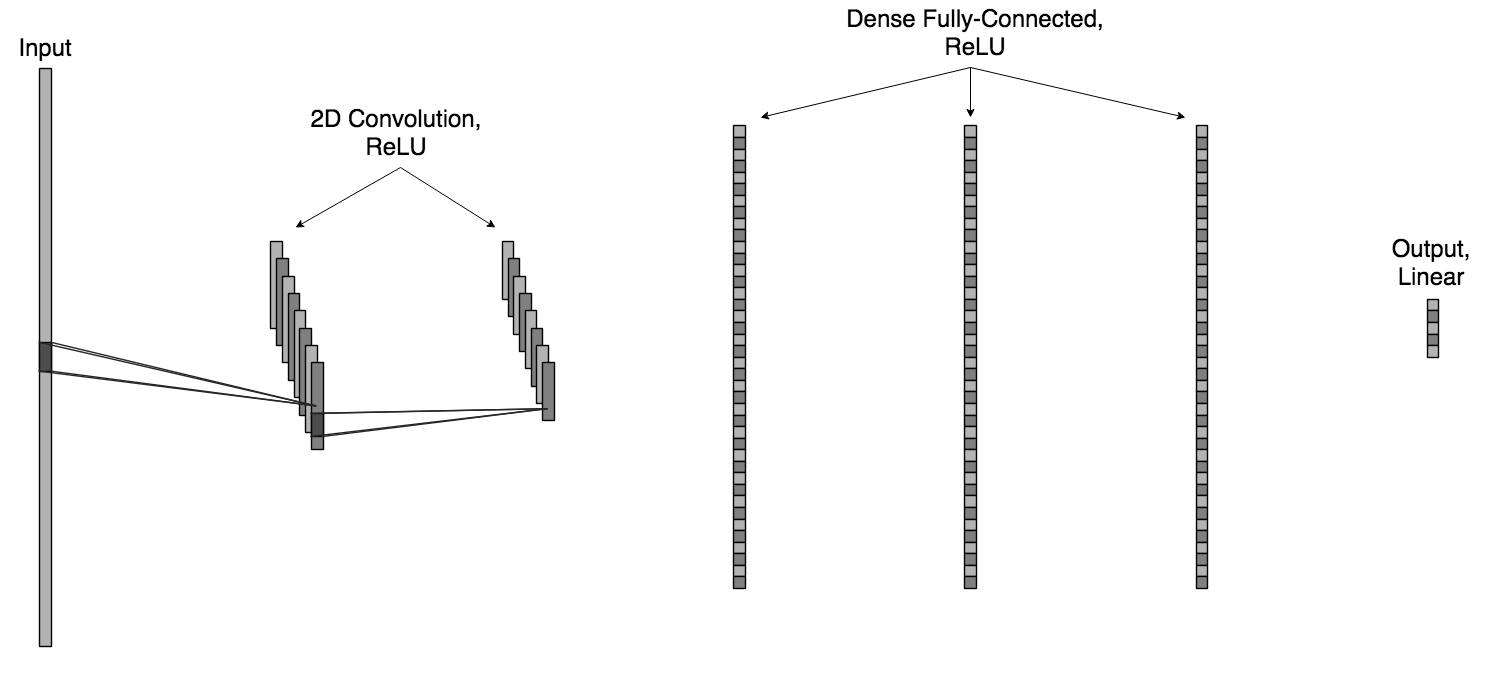}
	\caption{The CNN architecture designed for estimation of transmitter IQ imbalance.}
	\label{fig:model}
    \vspace{-15pt}
\end{figure}

The network architecture designed for the approach is shown in Figure \ref{fig:model}, and is loosely based off of the network architecture used in \cite{O’Shea2016}. 
To investigate the trade-offs between input size and performance, models were trained and evaluated using input sizes of 512, 1024, and 2048 raw IQ samples. 
Following the input layer, the network is composed of two two-dimensional convolutional layers and four dense fully-connected layers. 
Intuitively, the convolutional layers in this architecture are designed to identify and extract the relevant features, and the fully connected layers following are intended to perform the feature estimation \cite{5459469}. 

All layers, excluding the output layer, utilize a Rectified Linear Unit (ReLU) activation function. 
The ReLU function is a popular activation function in the literature, as it has been shown to be robust to saturation (when output is near zero or one), which usually causes learning to slow. 
However, because the function has a range of $[0, \infty)$, it cannot be used at the output layer, as it cannot produce negative estimates. 
Therefore, the final layer of the network uses the linear activation function to allow the network to estimate negative gain and phase imbalance values. 
The stochastic gradient descent algorithm used to train the networks was modified with a RMSProp optimizer and a mean squared error loss function \cite{rmsprop}.

Due to the complexity of estimating IQ imbalance, this approach estimates gain imbalance and phase imbalance separately using two different neural networks. 
Though both networks share the same underlying architecture (Figure \ref{fig:model}), using two networks allows each network to be optimized for the specific problem of gain imbalance estimation or phase imbalance estimation. 
The networks were trained using an iterative script which selected network parameters, then built and trained a
network with the given parameters, and evaluated the performance of the network. 
The parameters producing the network with the best performance after many iterations are then selected.
Therefore, the gain and phase imbalance estimation networks have different sized convolutional and dense layers, as well as different weights and biases, as they have been trained separately.
However, it should be noted that these two networks are not dependent upon each other, and therefore can be trained and run in parallel.

Similarly, separate networks were trained to estimate IQ imbalance for the simulated QAM and PSK signals. 
However, results in Section \ref{sec:est_results} will show that the performance of these networks is comparable, indicating the designed network architecture is not modulation specific. 
Additionally, though the networks have been trained per modulation type, they are generalizing over modulation order (i.e. the networks trained to estimate IQ imbalance for QAM can estimate gain and phase imbalances for QAM signals of orders 8, 16, 32, and 64). 

Each network used 2,020,000 sets of labeled samples in training: 2,000,000 sets of samples were used for training, 10,000 for validation, and 10,000 for model evaluation. 
The normalized mean squared error was used as the performance metric to determine the best network design and to evaluate performance, and is defined as
\begin{equation}
NMSE = \frac{1}{N} \sum_i \frac{(P_i - M_i)^2}{\overline{P}\hspace{2pt}\overline{M}}
\end{equation}
where $P$ is the vector of estimated offset values, $M$ is the vector of measured offset values, $\overline{P}$ is the mean of vector $P$, $\overline{M}$ is the mean of vector $M$, and $N$ is the length of vectors $P$ and $M$ \cite{nmse}. 

To further evaluate the performance of the estimators, evaluation sets were constructed with 180,000 sets of samples for the gain estimator and 200,000 sets of samples for the phase estimator. 
For each evaluation set, 1,000 sets of samples were generated at evenly spaced intervals of $\Delta \alpha = \pm 0.01$ for gain imbalance and evenly spaced intervals of $\Delta \theta = \pm 0.1^\circ$ for phase imbalance within the training range. 
These evaluation sets were used to determine the bias of the estimators and to generate the histograms shown in Section \ref{sec:sei_apprFit}.

\subsection{Simulation Results and Discussion}\label{sec:est_results}
\begin{figure}[t]
\vspace{-10pt}
  \centering
  \includegraphics[width=1.0\linewidth]{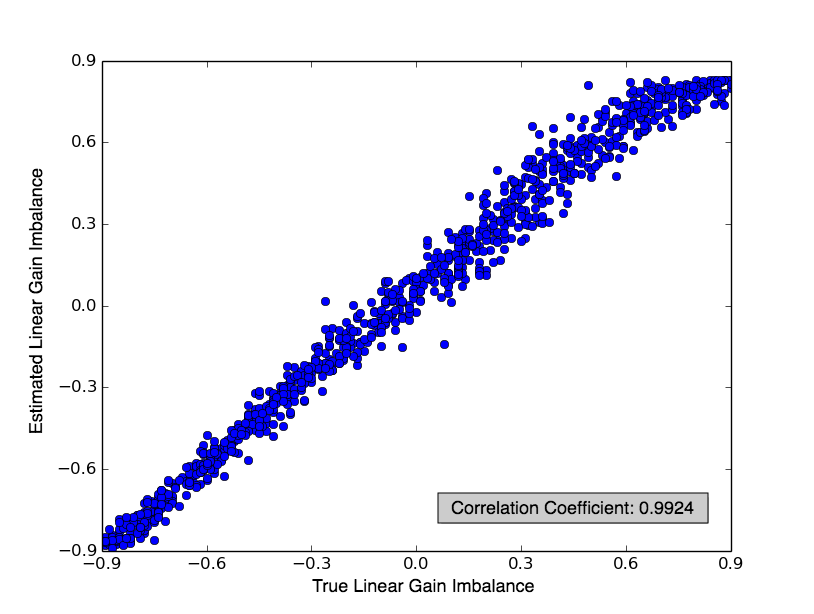}
  \caption{The true linear gain imbalance value versus the linear gain imbalance value estimated by the 1024-input CNN gain imbalance estimators with QAM input signals at 10dB SNR.}
  \label{fig:gainResults_QAM}
  \vspace{-10pt}
\end{figure}

\begin{figure}[t]
\vspace{-10pt}
  \centering
  \includegraphics[width=1.0\linewidth]{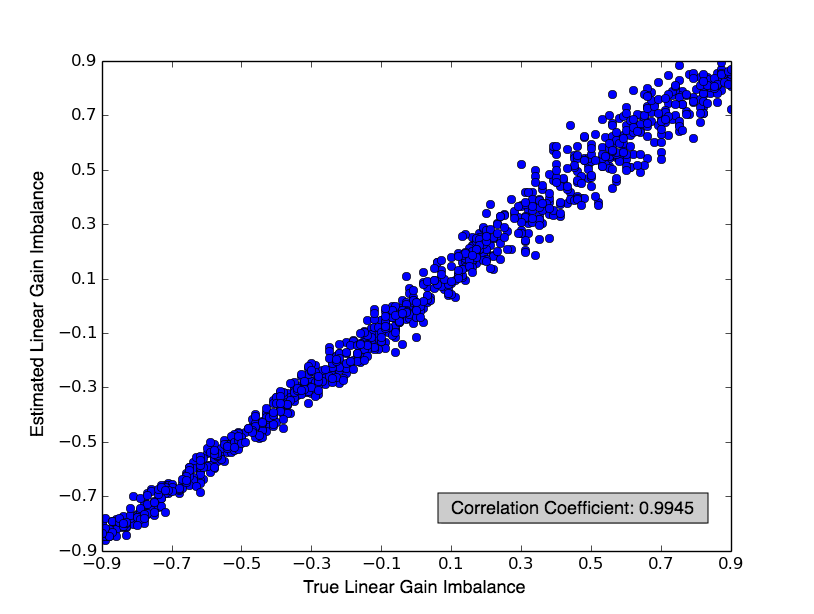}
  \caption{The true linear gain imbalance value versus the linear gain imbalance value estimated by the 1024-input CNN gain imbalance estimators with PSK input signals at 10dB SNR.}
  \label{fig:gainResults_PSK}
\end{figure}

\begin{figure}[t]
\vspace{-10pt}
  \centering
  \includegraphics[width=1.0\linewidth]{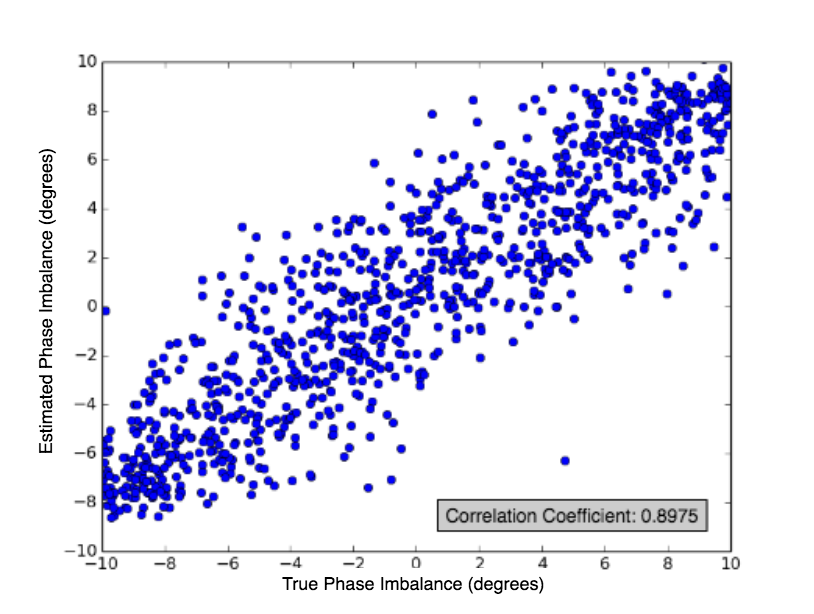}
  \caption{The true phase imbalance value versus the phase imbalance value estimated by the 1024-input CNN phase imbalance estimators with QAM input signals at 10dB SNR.}
  \label{fig:phaseResults_QAM}
  \vspace{-10pt}
\end{figure}

\begin{figure}[t]
\vspace{-10pt}
  \centering
  \includegraphics[width=1.0\linewidth]{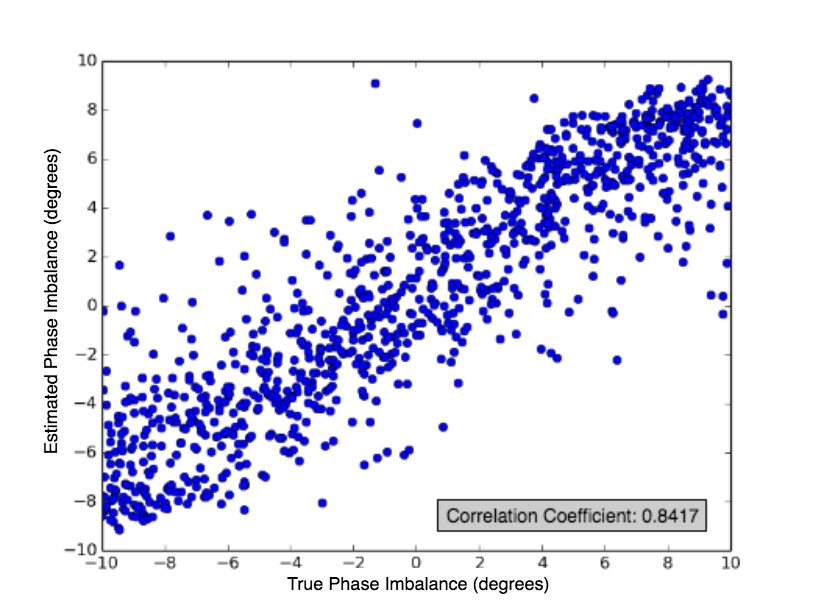}
  \caption{The true phase imbalance value versus the phase imbalance value estimated by the 1024-input CNN phase imbalance estimators with PSK input signals at 10dB SNR.}
  \label{fig:phaseResults_PSK}
\end{figure}

Initial results can be seen in Figures \ref{fig:gainResults_QAM}-\ref{fig:phaseResults_PSK}. 
The extremely strong linear correlations in Figure \ref{fig:gainResults_QAM} and \ref{fig:gainResults_PSK} shows each network's ability to estimate gain imbalance, for all offsets in the training range, using 1024 input samples. 
The phase imbalance estimators similarly show linear correlations in Figure \ref{fig:phaseResults_QAM} and \ref{fig:phaseResults_PSK}, though not nearly as strong as the gain imbalance estimators. 
This indicates phase imbalance is more difficult to estimate than gain imbalance, using the designed network architecture with 1024 input samples, as a strong linear correlation indicates a clear relationship between the estimated and true offset value, and far more input samples and/or a more sophisticated network is needed to estimate phase imbalance.


\subsubsection*{Estimator Bias}
\begin{figure}[t]
\vspace{-10pt}
  \centering
  \includegraphics[width=1.0\linewidth]{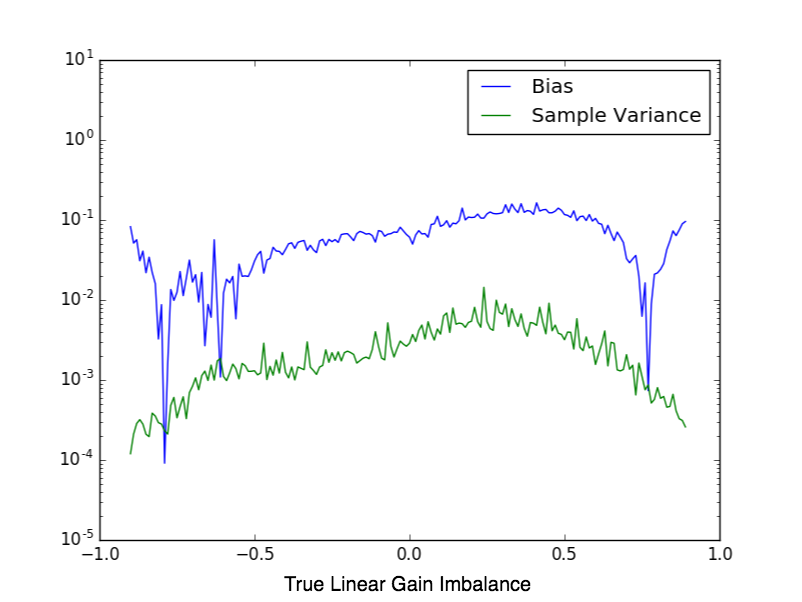}
  \caption{The bias and sample variance versus the true linear gain imbalance value for the 1024-input CNN gain imbalance estimator and QAM input signals simulated at 10dB SNR.}
  \label{fig:alphaBias_QAM}
  \vspace{-10pt}
\end{figure}

\begin{figure}[t]
\vspace{-10pt}
  \centering
  \includegraphics[width=1.0\linewidth]{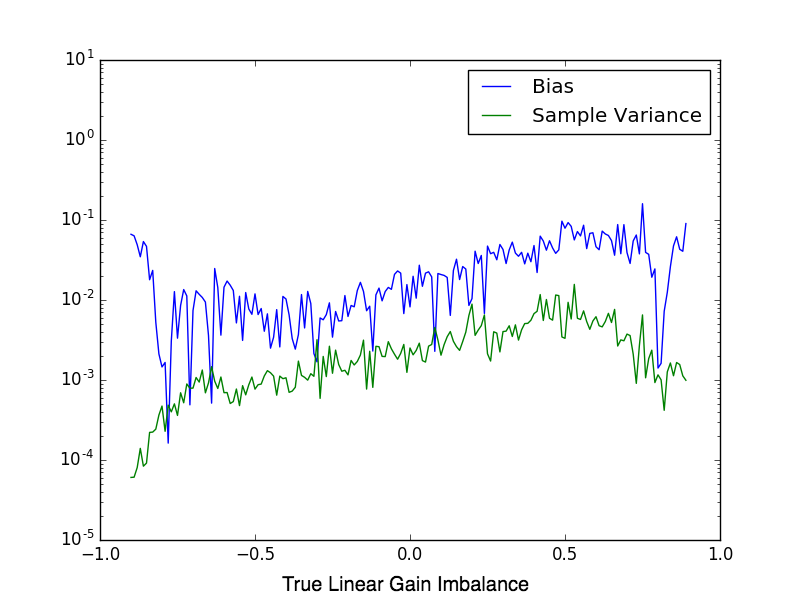}
  \caption{The bias and sample variance versus the true linear gain imbalance value for the 1024-input CNN gain imbalance estimator and PSK input signals simulated at 10dB SNR.}
  \label{fig:alphaBias_PSK}
\end{figure}

\begin{figure}[t]
\vspace{-10pt}
  \centering
  \includegraphics[width=1.0\linewidth]{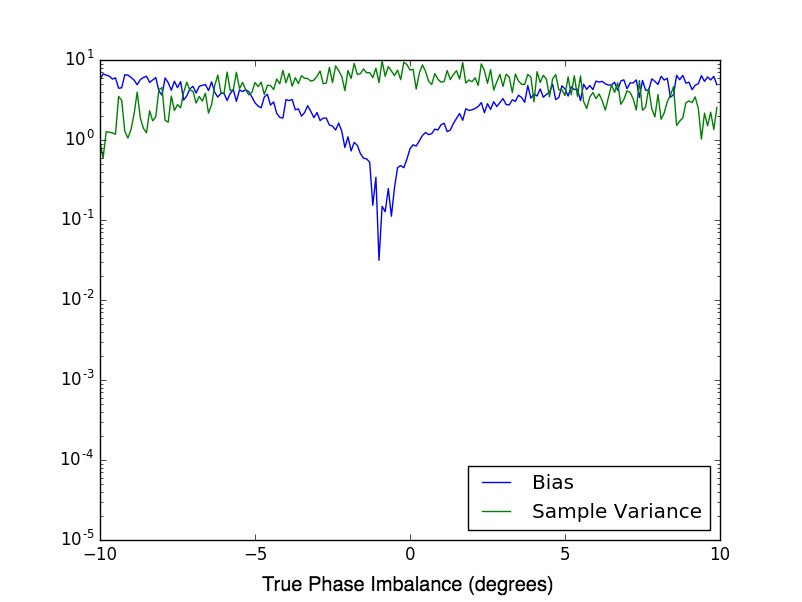}
  \caption{The bias and sample variance versus the true phase imbalance value for the 1024-input CNN phase imbalance estimator and QAM input signals simulated at 10dB SNR.}
  \label{fig:thetaBias_QAM}
  \vspace{-10pt}
\end{figure}

\begin{figure}[t]
\vspace{-10pt}
  \centering
  \includegraphics[width=1.0\linewidth]{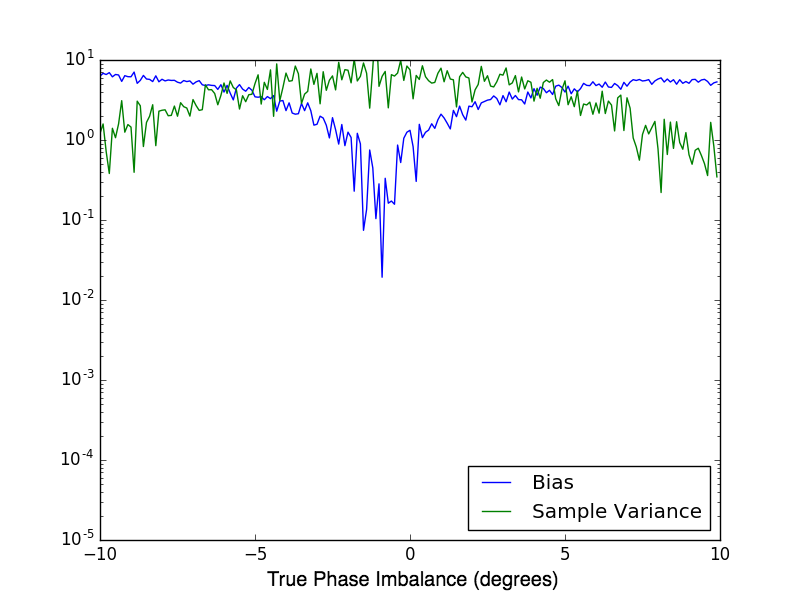}
  \caption{The bias and sample variance versus the true phase imbalance value for the 1024-input CNN phase imbalance estimator and PSK input signals simulated at 10dB SNR.}
  \label{fig:thetaBias_PSK}
\end{figure}

To examine the bias of the estimators, the cumulative average of the estimator outputs was taken for 1,000 sets of samples, each with the same offset value. 
The estimator can be called \emph{unbiased} if the cumulative moving average converges to the true offset value, and is \emph{biased} otherwise \cite{brown1947}. 
The results in Figures \ref{fig:alphaBias_QAM}-\ref{fig:thetaBias_PSK} show the bias and sample variance of the gain imbalance estimators and phase imbalance estimators respectively, as a function of the true offset value. 
The gain imbalance estimators produce estimates with low bias for all values within the training range ($-0.9$, $0.9$), with slightly higher bias values at the positive offset values. 
Additionally, the sample variance is also very low across all values within the training range. 
However, both the bias and the sample variance of the gain imbalance estimators are negligible in comparison to the bias and variance of the phase imbalance estimators, further indicating phase imbalance is far more difficult to estimate at 10dB SNR using this network architecture.



The phase imbalance estimators produce estimates with lowest bias when the true offset value is near zero. 
The bias then increases as the true offset value gets farther from zero in either direction. 
The sample variance shows an inverse trend, with maximum sample variance near zero and minimum sample variance at $-10^\circ$ and $10^\circ$. 
These trends further emphasize the inaccuracy of the phase imbalance estimators across all values. 

\subsubsection*{Impact of SNR and Network Input Size on Performance}
\begin{figure}[t]
\vspace{-10pt}
  \centering
  \includegraphics[width=1.0\linewidth]{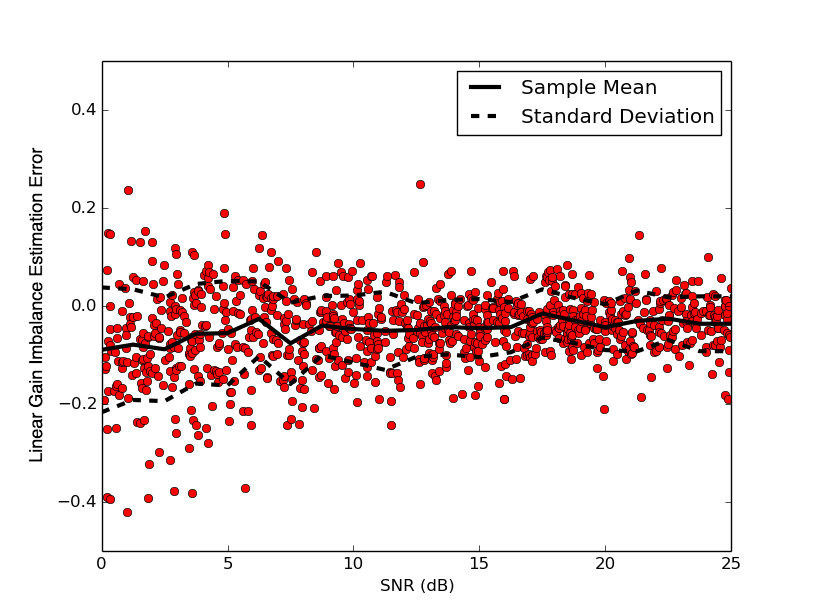}
  \caption[The Linear Gain Imbalance Estimation Errors for QAM signals simulated with SNRs between 0dB and 25dB.]{The Linear Gain Imbalance Estimation Errors for signals simulated with SNRs between 0dB and 25dB. True linear gain imbalances vary uniformly between [-0.9, 0.9].}
  \label{fig:gainSNR_QAM}
  \vspace{-10pt}
\end{figure}

\begin{figure}[t]
\vspace{-10pt}
  \centering
  \includegraphics[width=1.0\linewidth]{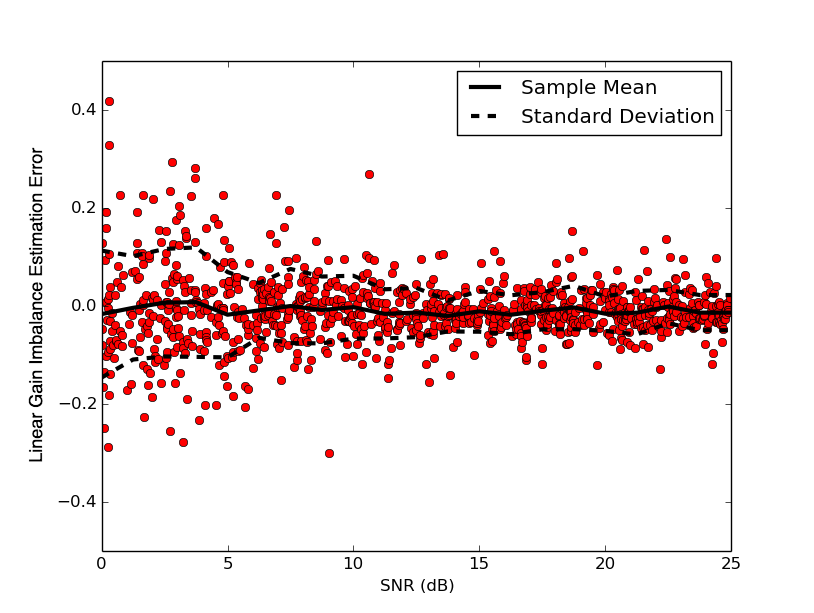}
  \caption[The Linear Gain Imbalance Estimation Errors for PSK signals simulated with SNRs between 0dB and 25dB.]{The Linear Gain Imbalance Estimation Errors for signals simulated with SNRs between 0dB and 25dB. True linear gain imbalances vary uniformly between [-0.9, 0.9].}
  \label{fig:gainSNR_PSK}
\end{figure}

\begin{figure}[t]
\vspace{-10pt}
  \centering
  \includegraphics[width=1.0\linewidth]{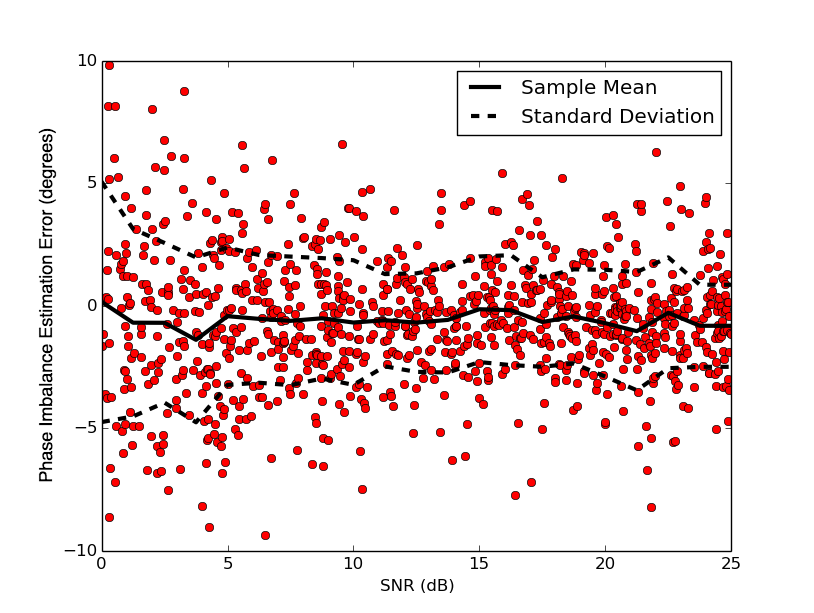}
  \caption[The Phase Imbalance Estimation Errors for QAM signals simulated with SNRs between 0dB and 25dB.]{The Phase Imbalance Estimation Errors for signals simulated with SNRs between 0dB and 25dB. True phase imbalances vary uniformly between [-10$^\circ$, 10$^\circ$].}
  \label{fig:phaseSNR_QAM}
  \vspace{-10pt}
\end{figure}

\begin{figure}[t]
\vspace{-10pt}
  \centering
  \includegraphics[width=1.0\linewidth]{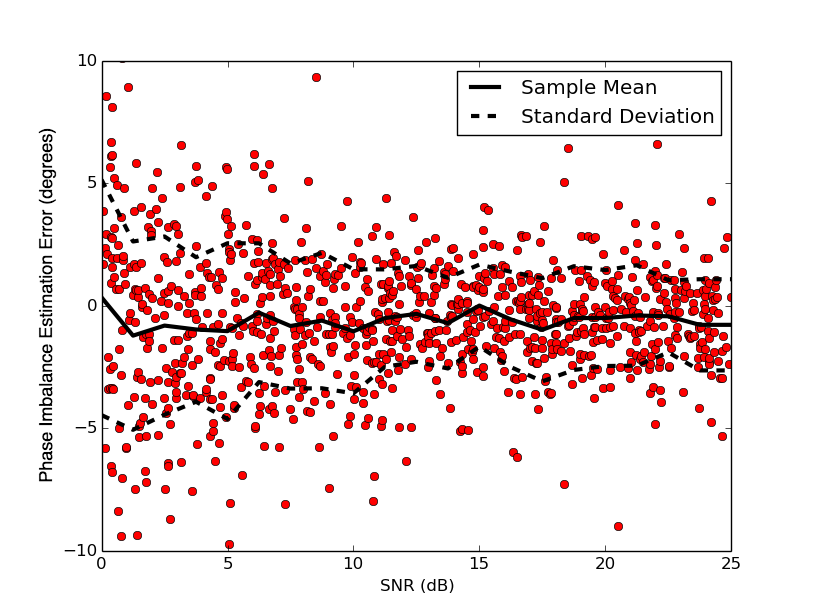}
  \caption[The Phase Imbalance Estimation Errors for PSK signals simulated with SNRs between 0dB and 25dB.]{The Phase Imbalance Estimation Errors for signals simulated with SNRs between 0dB and 25dB. True phase imbalances vary uniformly between [-10$^\circ$, 10$^\circ$].}
  \label{fig:phaseSNR_PSK}
\end{figure}

The effect of SNR on the performance of the estimators can be seen in Figures \ref{fig:gainSNR_QAM}-\ref{fig:phaseSNR_PSK}. 
For both offset estimators trained for both modulation types, it is shown that as the SNR increases, the Linear Gain Imbalance Estimation Error (the difference between the true and estimated imbalances) decreases. 
However, the mean imbalance error stays almost constant near zero for all SNR values with the standard deviation of the offset error decreasing as SNR increases, with diminishing returns after 10dB.


\begin{figure}[t]
\vspace{-10pt}
  \centering
  \includegraphics[width=1.0\linewidth]{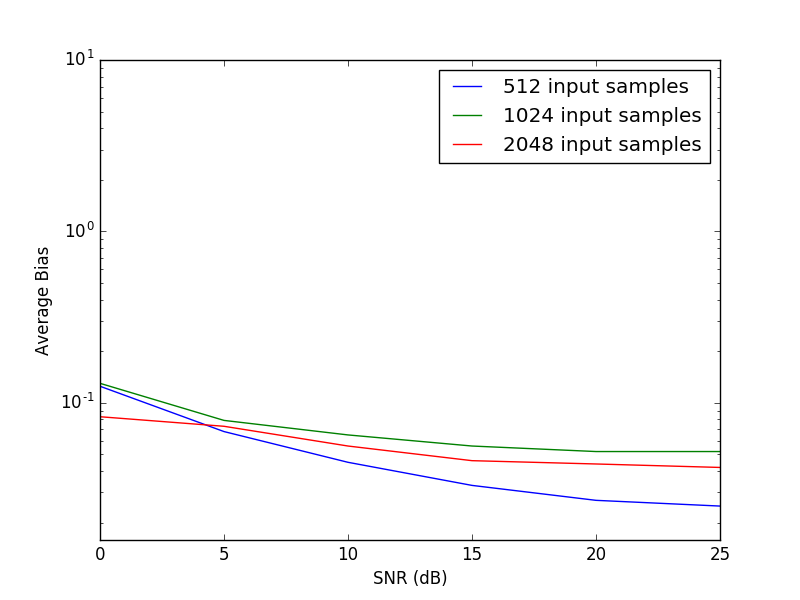}
  \caption{The average bias versus SNR for CNN gain imbalance estimators, trained on QAM signals, with input sizes of 512 samples, 1024 samples, and 2048 samples.}
  \label{fig:alphaBiasVsnr_QAM}
  \vspace{-10pt}
\end{figure}

\begin{figure}[t]
\vspace{-10pt}
  \centering
  \includegraphics[width=1.0\linewidth]{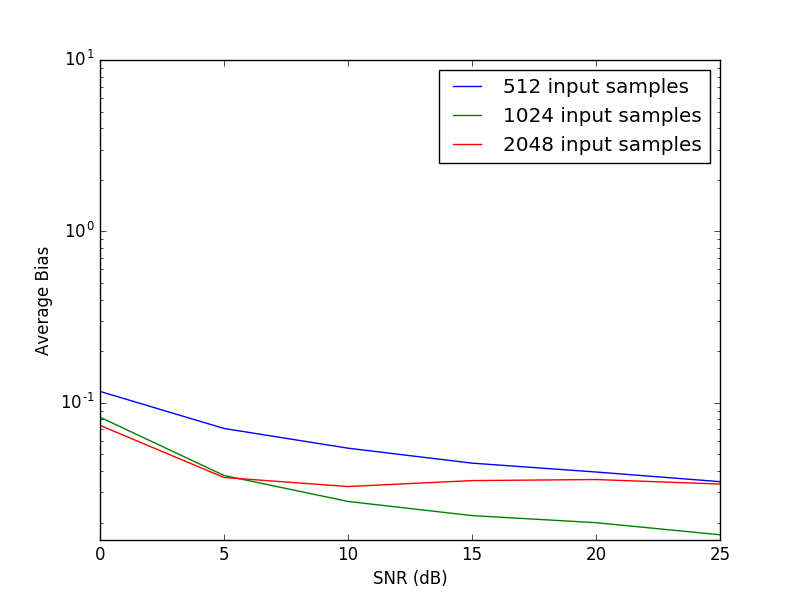}
  \caption{The average bias versus SNR for CNN gain imbalance estimators, trained on PSK signals, with input sizes of 512 samples, 1024 samples, and 2048 samples.}
  \label{fig:alphaBiasVsnr_PSK}
\end{figure}

\begin{figure}[t]
\vspace{-10pt}
  \centering
  \includegraphics[width=1.0\linewidth]{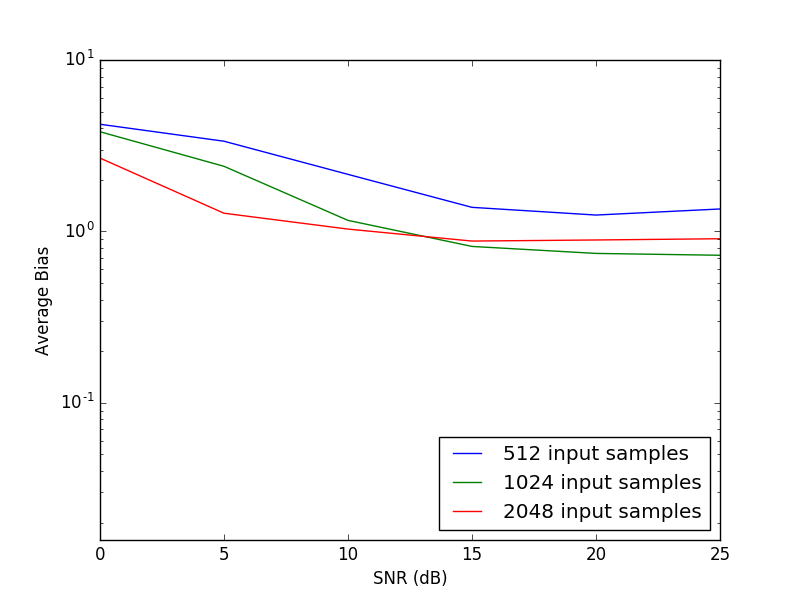}
  \caption{The average bias versus SNR for CNN phase imbalance estimators, trained on QAM signals, with input sizes of 512 samples, 1024 samples, and 2048 samples.}
  \label{fig:thetaBiasVsnr_QAM}
  \vspace{-10pt}
\end{figure}

\begin{figure}[t]
\vspace{-10pt}
  \centering
  \includegraphics[width=1.0\linewidth]{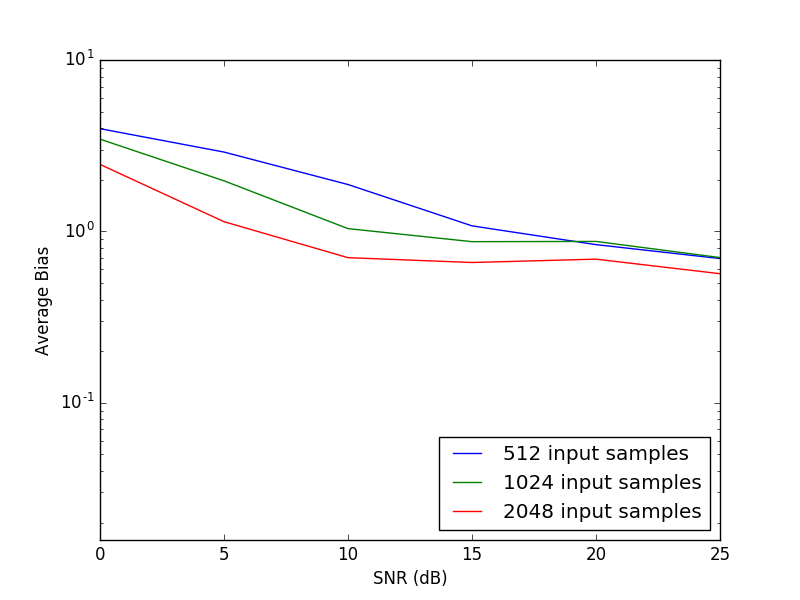}
  \caption{The average bias versus SNR for CNN phase imbalance estimators, trained on PSK signals, with input sizes of 512 samples, 1024 samples, and 2048 samples.}
  \label{fig:thetaBiasVsnr_PSK}
\end{figure}

\begin{figure}[t]
\vspace{-10pt}
  \centering
  \includegraphics[width=1.0\linewidth]{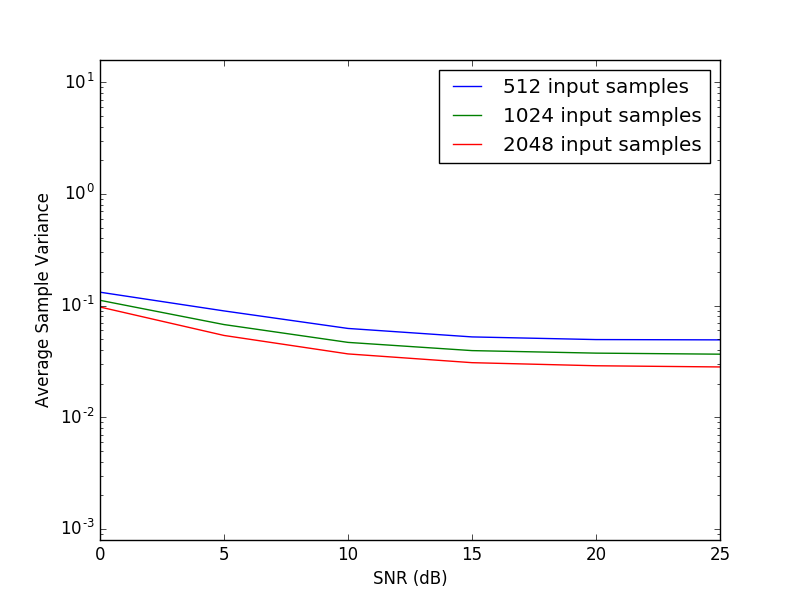}
  \caption{The sample variance of the histograms for the 512-input, 1024-input, and 2048-input CNN gain imbalance estimators, trained on QAM signals, as a function of SNR.}
  \label{fig:gainSV_QAM}
  \vspace{-10pt}
\end{figure}

\begin{figure}[t]
\vspace{-10pt}
  \centering
  \includegraphics[width=1.0\linewidth]{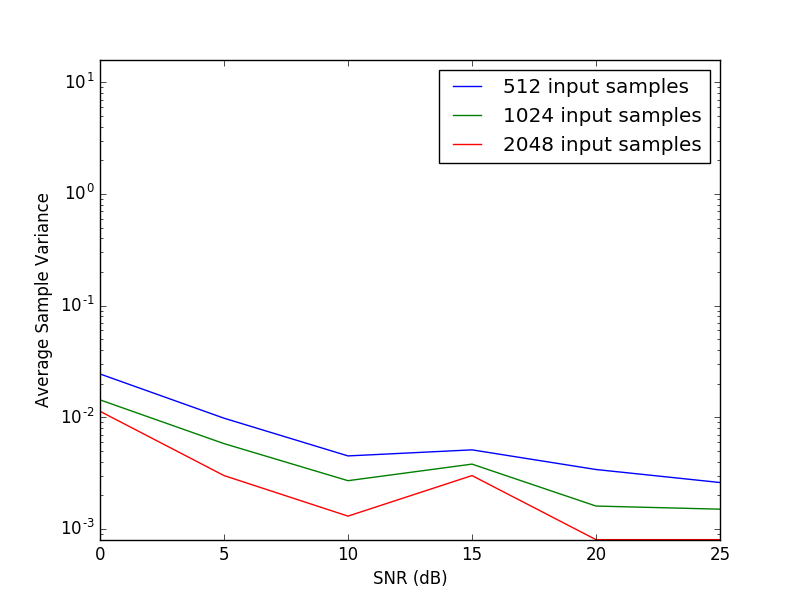}
  \caption{The sample variance of the histograms for the 512-input, 1024-input, and 2048-input CNN gain imbalance estimators, trained on PSK signals, as a function of SNR.}
  \label{fig:gainSV_PSK}
\end{figure}

\begin{figure}[t]
\vspace{-10pt}
  \centering
  \includegraphics[width=1.0\linewidth]{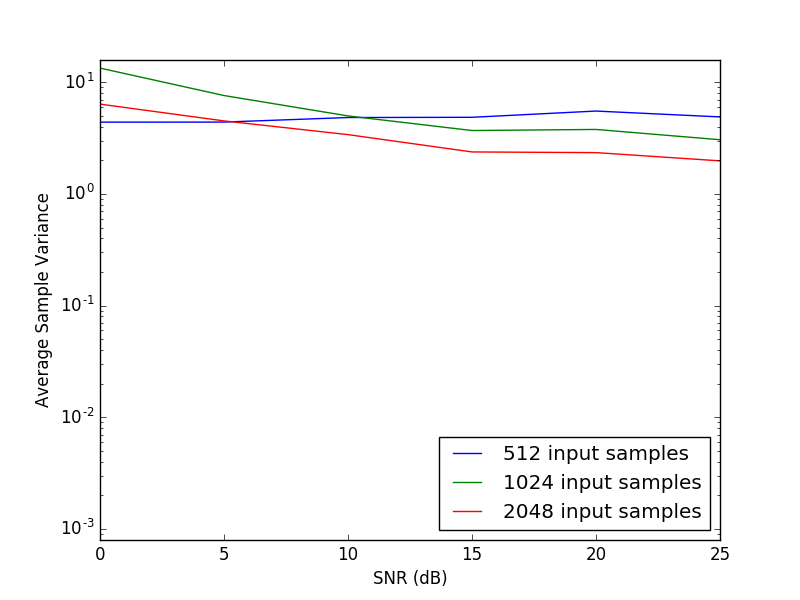}
  \caption{The sample variance of the histograms for the 512-input, 1024-input, and 2048-input CNN phase imbalance estimators, trained on QAM signals, as a function of SNR.}
  \label{fig:phaseSV_QAM}
  \vspace{-10pt}
\end{figure}

\begin{figure}[t]
\vspace{-10pt}
  \centering
  \includegraphics[width=1.0\linewidth]{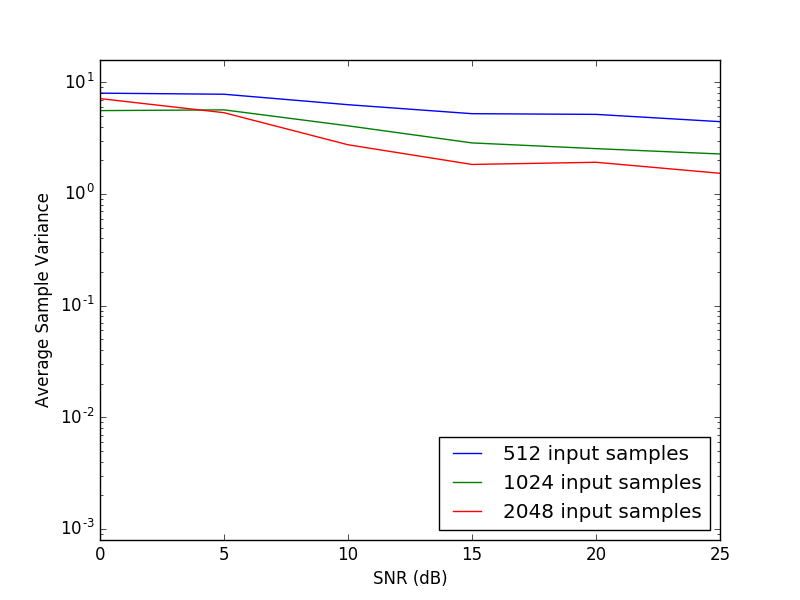}
  \caption{The sample variance of the histograms for the 512-input, 1024-input, and 2048-input CNN phase imbalance estimators, trained on PSK signals, as a function of SNR.}
  \label{fig:phaseSV_PSK}
  \vspace{-10pt}
\end{figure}

The effect of the network input size and the SNR of the input signal on the performance of the gain and phase imbalance estimators was further investigated using the average bias and the sample variance of the output. 
These results are shown in Figures \ref{fig:alphaBiasVsnr_QAM}-\ref{fig:phaseSV_PSK}. 
In Figures \ref{fig:alphaBiasVsnr_QAM}-\ref{fig:thetaBiasVsnr_PSK}, as the SNR increases, the bias of the estimators decreases. 
Additionally, in Figures \ref{fig:gainSV_QAM} and \ref{fig:gainSV_PSK}, for the gain imbalance estimators, as the SNR of the input signal increases, the sample variance also decreases, with dramatic improvement between $0-10$dB and diminishing returns after $15$dB. 
Shown in Figures \ref{fig:phaseSV_QAM} and \ref{fig:phaseSV_PSK}, the PSK phase imbalance estimators and the 1024- and 2048-input QAM phase imbalance estimators behave similarly. 
However, the sample variance of the 512-input QAM phase imbalance estimator histogram remains constant for all SNRs. 
This, in addition to the high average bias of the 512-input QAM phase imbalance estimator, suggests that 512 input samples does not give the network enough information to learn phase imbalance for the QAM modulation type. 
Therefore the network produces very similar outputs for all inputs at all SNRs.
The results in Figures \ref{fig:gainSV_QAM}-\ref{fig:phaseSV_PSK} also show, as the number of input samples increases, the sample variance decreases, excluding the 512-input QAM phase imbalance estimator. 
This behavior is expected, as with more input samples, the network sees the signal for longer, and therefore has more information about the signal to use in its estimation.




From the results shown above, it can be concluded that though increasing the number of input samples to the network may not increase the accuracy of the estimate, the network does become more confident in the estimate it produces. 
However, it should be noted that though more inputs generally improves some aspects of performance, it also slows the network and increases training time, as it has to process much more information. 
Additionally, increasing the input size to the network also requires more training data, as $n$ sets of $1024$ samples requires twice the memory as $n$ sets of $512$ samples.
Given the results shown above, $1024$ samples provides good accuracy without requiring prohibitively large training sets and slow training times. 

\section{Transmitter Gain Imbalance Estimation for SEI}\label{sec:estSEI}
\begin{figure}[t]
\vspace{10pt}
	\centering
	\includegraphics[width=1.0\linewidth]{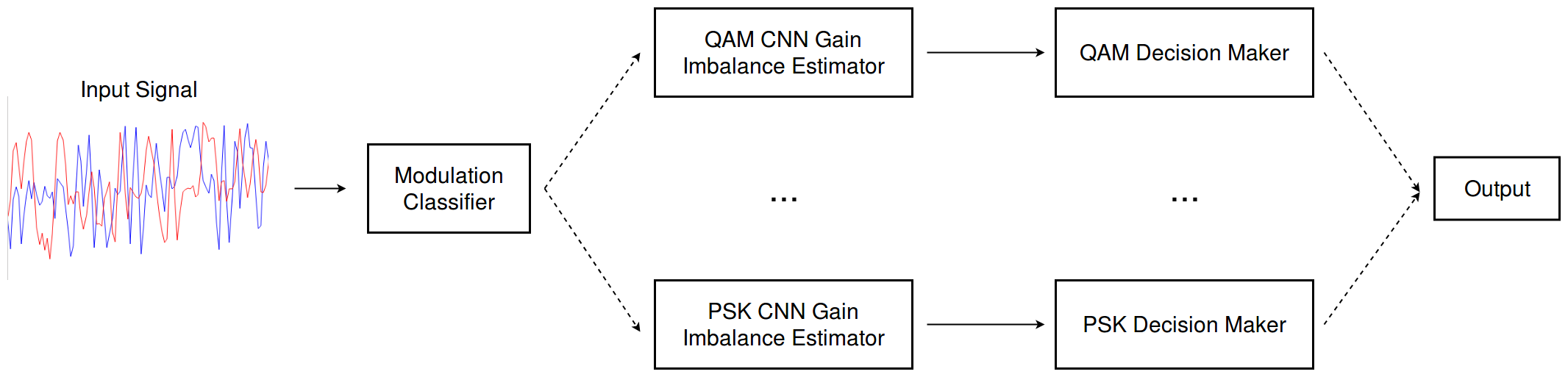}
	\caption{The designed emitter identification approach using CNN IQ imbalance estimators.}
	\label{fig:sei_approach}
    \vspace{-10pt}
\end{figure}

Three main steps are required to perform emitter identification using the proposed approach, shown in Figure \ref{fig:sei_approach}: modulation classification, gain imbalance estimation, and decision making. The result is a decision tree-like structure in which the output of each step informs the next action, as described below. 

The first step is modulation classification because the pre-trained CNN gain imbalance estimators are modulation-specific.
It is important to note that while any modulation classifier may be used, a key advantage of the developed approach over traditional approaches is the use of only the raw IQ as input.
In order to retain this advantage, the modulation classifier should only use raw IQ as input as well.
Such modulation classifiers exist in the literature.
For example, \cite{steve_milcom} uses a CNN architecture to perform modulation classification using only raw IQ as input.

The output of the modulation classifier determines which modulation-specific CNN gain imbalance estimator the input signal is fed to, and the gain imbalance of the emitter can be appropriately estimated. 
The point estimate produced by the CNN gain imbalance estimator in the previous step is then used to determine the identity of the transmitter using modulation-specific decision makers built using Gaussian probability density functions (\emph{pdf}s) and Bayes optimal decision boundaries, to be discussed.

\subsection{Gaussian Curve Fit to CNN Output Histograms}\label{sec:sei_apprFit}
\begin{figure}[t]
\vspace{-10pt}
  \centering
  \includegraphics[width=1.0\linewidth]{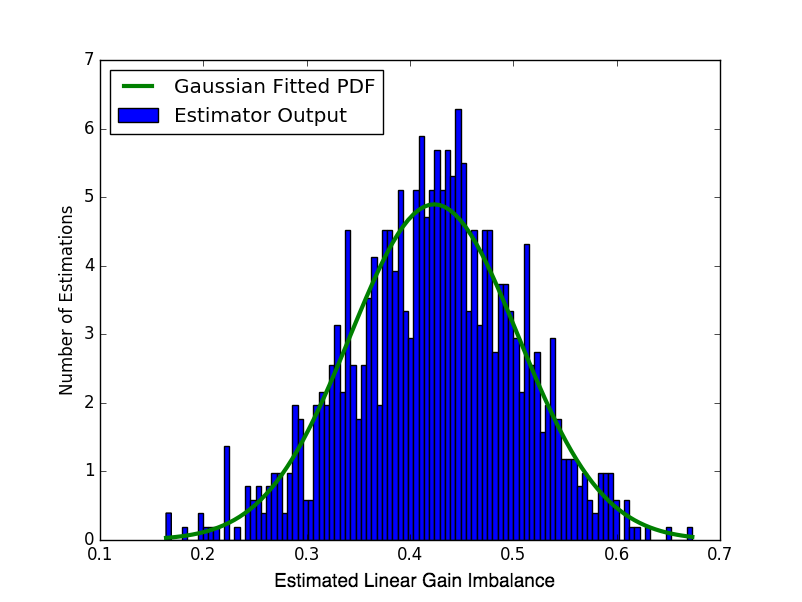}
  \caption{The fitted Gaussian curve for the 1024 input CNN gain imbalance estimator output histogram with QAM input signals at 10dB SNR. (Linear gain imbalance = 0.30)}
  \label{fig:gainGOF_QAM}
\end{figure}

\begin{figure}[t]
\vspace{-10pt}
  \centering
  \includegraphics[width=1.0\linewidth]{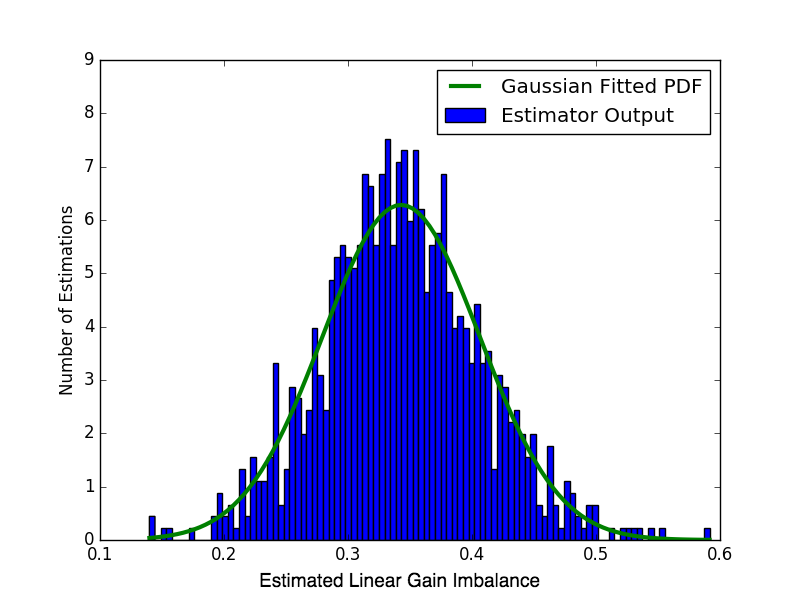}
  \caption{The fitted Gaussian curve for the 1024 input CNN gain imbalance estimator output histogram with PSK input signals at 10dB SNR. (Linear gain imbalance = 0.30)}
  \label{fig:gainGOF_PSK}
  \vspace{-10pt}
\end{figure}

Using the evaluation sets described in Section \ref{sec:est_model}, histograms can be produced for the CNN estimator outputs at evenly spaced intervals of 0.01 within the training interval, [-0.9, 0.9]. 
Because a Gaussian trend was observed, the \emph{pdf} was fitted to the gain imbalance estimator output histograms, as shown in Figures \ref{fig:gainGOF_QAM} and \ref{fig:gainGOF_PSK}.


The goodness of fit was tested using the Chi-Squared Goodness of Fit (GoF) test, as follows \cite{Pearson1992}: letting the null hypothesis ($H_0$) be that the data is consistent with a Gaussian distribution, and the alternate hypothesis ($H_1$) be the data is not consistent with a Gaussian distribution, the $\chi^2$ test produces a $p$-value representing the probability of incorrectly rejecting the null hypothesis. 
The null hypothesis is rejected if the $p$-value is less than the chosen significance level. 
In the literature, 0.05 is a commonly chosen significance level and is used here \cite{Pearson1992, chi_guide}. As shown in Table \ref{table:gof_pvals}, the $\chi^2$ test produced average $p$-values greater than $0.05$ for both the QAM and PSK gain imbalance estimators for SNRs varying from 0 to 25dB, over all offset values, using 1024 input samples, so the null hypothesis, and thus the Gaussian fit for the CNN outputs, was accepted.

\begin{table}[t]
\vspace{-10pt}
\centering
\caption{The average $p$-values produced by the $\chi^2$ GoF test.}
\begin{tabular}{@{}l|ll@{}}
\toprule
 & QAM & PSK \\ \midrule
0dB & 0.519 & 0.717 \\
5dB & 0.644 & 0.565 \\
10dB & 0.707 & 0.710 \\
15dB & 0.659 & 0.600 \\
20dB & 0.591 & 0.558 \\
25dB & 0.618 & 0.525 \\ \bottomrule
\end{tabular}
\label{table:gof_pvals}
\end{table}

\subsection{Bayesian Decision Boundaries}\label{sec:bayes}
Given two offset values, $i$ and $j$, the Bayesian decision boundary between the fitted \emph{pdf}s, $p(x \vert i)$ and $p(x \vert j)$, is calculated as follows \cite{Duda}. The following calculations assume that any given emitter is equally likely to have any gain imbalance value, but are not specific to the Gaussian \emph{pdf} and can therefore be used for any curve fit. 

\begin{figure}[t]
\vspace{-10pt}
	\centering
	\includegraphics[width=1.0\linewidth]{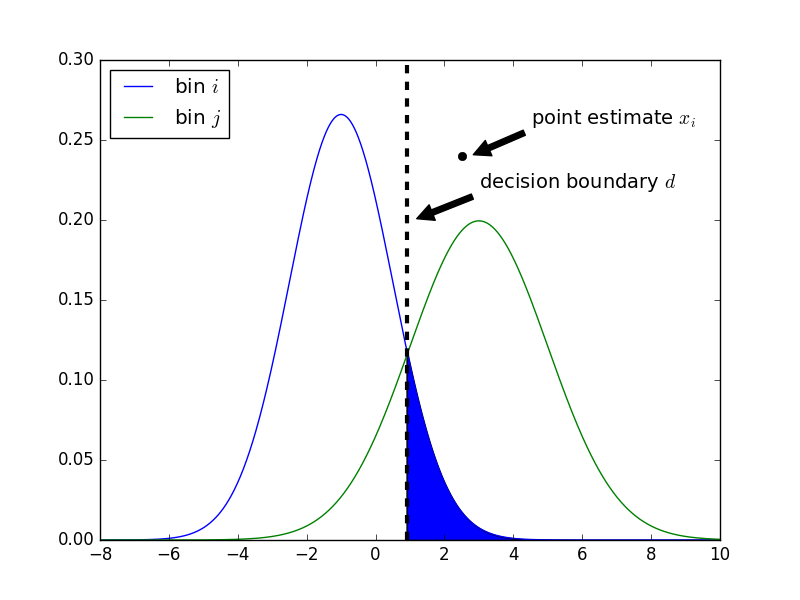}
	\caption{An example decision scenario.}
	\label{fig:decision}
    \vspace{-10pt}
\end{figure}

Letting $x$ be the received signal data, 
\begin{align*}
\textrm{Decide $i$ if $P(i \vert x) > P(j \vert x)$; otherwise decide $j$}.
\end{align*}
Using Bayes Rule, this decision rule can be expressed in terms of the fitted \emph{pdf}s ($p(x \vert i), p(x \vert j)$) and the probability of the emitter have a given gain imbalance value ($P(i), P(j)$):
\begin{align*}
\textrm{Decide $i$ if $p(x \vert i)P(i) > p(x \vert j)P(j)$; otherwise decide $j$}.
\end{align*}
Finally, assuming each gain imbalance value is equally likely to occur (i.e. $P(i) = P(j)$), the final decision rule is
\begin{align*}
\textrm{Decide $i$ if $p(x \vert i) > p(x \vert j)$; otherwise decide $j$},
\end{align*}
making the decision boundary the intersection point(s) of the two fitted \emph{pdf}s $p(x \vert i)$ and $p(x \vert j)$ for gain imbalance values $i$ and $j$. More specifically, the decision boundary is $d = p(x \vert i) = p(x \vert j)$, as shown in Figure \ref{fig:decision}.

\subsection{Decision Making}\label{sec:decision}
Given a decision boundary, $d$, calculated between the two \emph{pdf}s for gain imbalance values $i$ and $j$ and a received signal, a decision can be made about emitter identity. After modulation classification, the received signal can be fed to the appropriate gain imbalance estimator, producing a point estimate, $x_i$, of the gain imbalance of the emitter which sent the signal. 

Without loss of generality, let the mean of $p(x \vert i)$ be less than the mean of $p(x \vert j)$, as shown in the example decision scenario in in Figure \ref{fig:decision}. If the point estimate falls on the left side of the decision boundary, it is decided the transmitted signal came from an emitter with gain imbalance value $i$. Otherwise, it is decided the transmitted signal came from an emitter with gain imbalance value $j$. 
In the case that the emitters in the system are known, the \emph{pdf}s of the known gain imbalance values can be selected and the decision boundaries between these \emph{pdf}s calculated. Decisions on point estimates are then made as described above. While this method has use cases for Dynamic Spectrum Access and cooperative scenarios \cite{Song2012}, the ability to perform SEI in non-cooperative and blind scenarios is a primary motivator of this work.
In the case that the emitters in the system are not known, the approach may still be used. However, because the \emph{pdf}s of the known gain imbalance values cannot be selected, it is only possible to bin the emitters by their gain imbalance value. To do this, \emph{pdf}s are selected at evenly spaced values and the decision boundaries are calculated. Then, as in the first case, decisions on point estimates are made as described above. For simplicity, the results shown in Section \ref{sec:results} will consider only this second case.  

\subsection{The Probability of Mis-Identifying Emitters}\label{sec:misid}
Given two fitted \emph{pdf}s, $p(x \vert i)$ and $p(x \vert j)$, and the optimal decision boundary, $d$, between the \emph{pdf}s, it is also possible to determine the probability of incorrectly identifying an emitter.
Consider the scenario in Figure \ref{fig:decision}, where a point estimate, $x_i$, is produced from a set of samples received from an emitter belonging to gain imbalance bin $i$. Again, without loss of generality, let the mean of $p(x \vert i)$ be less than the mean of $p(x \vert j)$. A correct classification occurs when the estimate from the CNN, $x_i$, is less than the decision boundary $d$. Therefore, an incorrect classification occurs when $x_i > d$. Because the area under a \emph{pdf} is $1$, the probability of this occurring is represented by 
\begin{align*}
\int_d^\infty p(x \vert i) dx.
\end{align*}

\subsection{Model Design, Training, and Evaluation}\label{sec:model}
\begin{figure}[t]
\vspace{-10pt}
	\centering
	\includegraphics[width=1.0\linewidth]{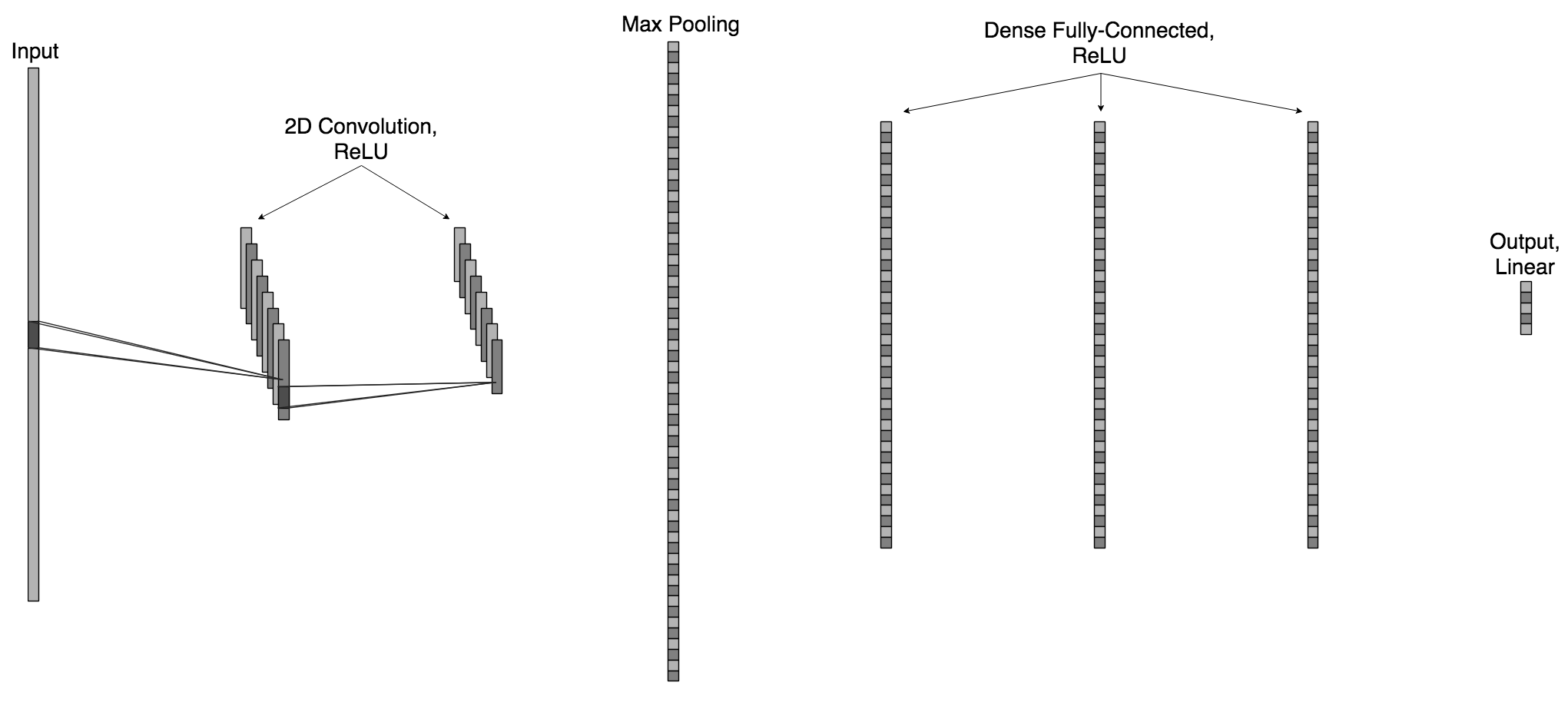}
	\caption{The CNN architecture designed for estimation of transmitter gain imbalance to perform SEI.}
	\label{fig:model_SEI}
    \vspace{-10pt}
\end{figure}

The model developed previously contained two two-dimensional convolutional layers followed by four dense fully-connected layers. 
The final layer used a linear activation function, while all other layers used the ReLU activation function. 
This approach uses the same model, modified with one max-pooling layer, with size $= 2$, inserted between the convolutional layers and the dense layers, as shown in Figure \ref{fig:model_SEI}. 

When determining which networks performed best, the NMSE was no longer a helpful evaluation metric, as a network with a low average NMSE could produce histograms with larger variance than networks with a higher average NMSE. 
Therefore, to evaluate the performance of trained networks, the evaluation sets previously described in Section \ref{sec:est_model} were used. 
For a given trained network, \emph{pdf}s were fitted for each of the gain imbalance values, and the minimum gain imbalance separation needed to obtain a probability of mis-identification of less than $5\%$ was calculated. 
This minimum gain imbalance separation value was used to determine which networks were performing better than others. 

\subsection{Simulation Results and Discussion}\label{sec:results} 
To fully analyze the performance of the designed SEI approach, the following sections investigate the effect of SNR, true gain imbalance value, and modulation scheme on the SEI decision, and compare the designed approach to a traditional feature-based approach. Further, the practicality of using the designed approach is discussed, considering the IQ imbalance values typically found in real systems and the assumptions typically made by traditional SEI systems. 

\subsubsection*{Impact of SNR on SEI Performance}
\begin{figure}[t]
  \centering
  \includegraphics[width=1.0\linewidth]{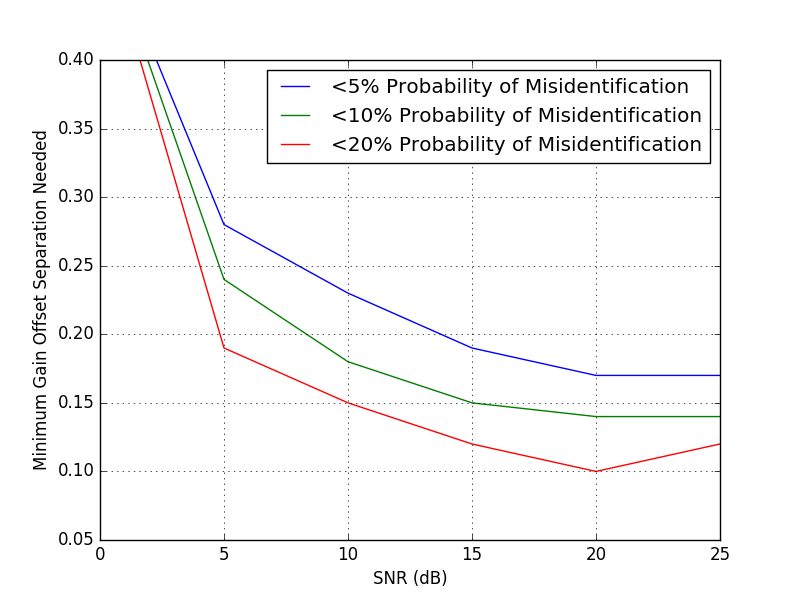}
  \caption{The SNR vs minimum gain imbalance separation needed to obtain $<5\%$, $<10\%$, and $<20\%$ probability of mis-identification using the CNN gain imbalance estimator trained on QAM input signals.}
  \label{fig:snr_QAM}
  \vspace{-10pt}
\end{figure}

\begin{figure}[t]
\vspace{-10pt}
  \centering
  \includegraphics[width=1.0\linewidth]{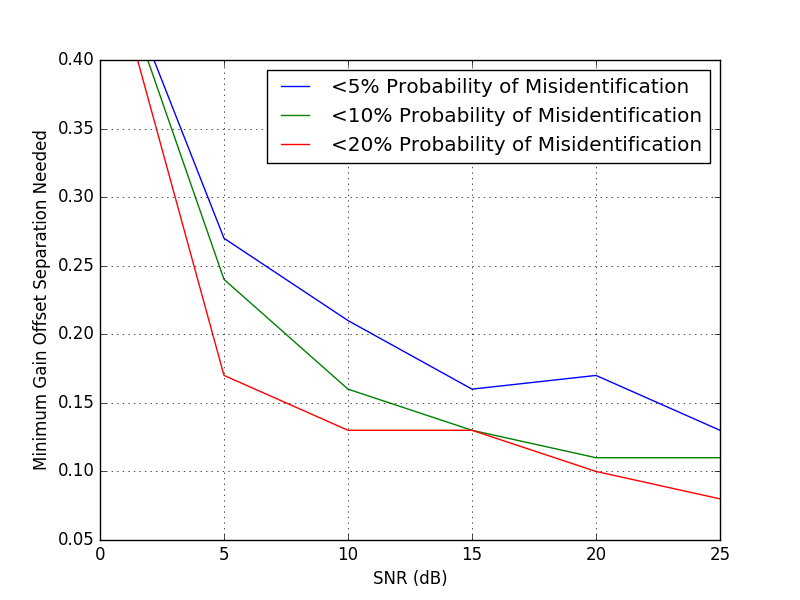}
  \caption{The SNR vs minimum gain imbalance separation needed to obtain $<5\%$, $<10\%$, and $<20\%$ probability of mis-identification using the CNN gain imbalance estimator trained on PSK input signals.}
  \label{fig:snr_PSK}
  \vspace{-10pt}
\end{figure}

Using the evaluation sets constructed for QAM and PSK, the minimum gain imbalance separations needed to obtain average probabilities of mis-identification of less $20\%$, $10\%$, and $5\%$ across all gain imbalance values were calculated, as described in Section \ref{sec:misid}. 
The impact of SNR on the ability to identify emitters at these levels of accuracy is shown in Figures \ref{fig:snr_QAM} and \ref{fig:snr_PSK}. 
At $0dB$, the estimators cannot be used to perform emitter identification to even a $80\%$ probability of correct identification. 
However, as the SNR increases, the minimum gain imbalance separation needed to obtain $<5\%$, $<10\%$, and $<20\%$ probabilities of mis-identification decreases with diminishing returns at around $20dB$. 
Therefore, the lower the probability of mis-identification needed in a system, the higher the gain imbalance separation needed.


\subsubsection*{Impact of Gain Imbalance Value on SEI Ability}

In Section \ref{sec:est_results} and Figures \ref{fig:alphaBias_QAM} and \ref{fig:alphaBias_PSK}, it was shown that the sample variance of the gain imbalance estimators is slightly lower when the true offset value is at the limits of the training range (near $-0.9$ and $0.9$). 
As a result, the variance of the fitted \emph{pdf}s is lower when the true gain imbalance value is near the limits of the training range, in comparison to when the true gain imbalance value is near zero. 
Therefore, the probability of mis-identification is lower when the true gain imbalance value is near the limits of the training range.

\subsubsection*{Impact of Modulation Scheme on SEI Performance}

The ability to perform gain imbalance estimation on QAM and PSK signals using the designed CNN architecture was shown in Section \ref{sec:est_results}. 
Though the use of CNN estimators for gain imbalance estimation on other signal types was not investigated, the comparable results of the CNN gain imbalance estimators trained for QAM and PSK showed that the designed network architecture described in Section \ref{sec:est_model} was not modulation specific. 
Investigation into the performance of the estimators on further modulation schemes is left for future work.

\begin{figure}[t]
\vspace{-10pt}
	\centering
	\includegraphics[width=1.0\linewidth]{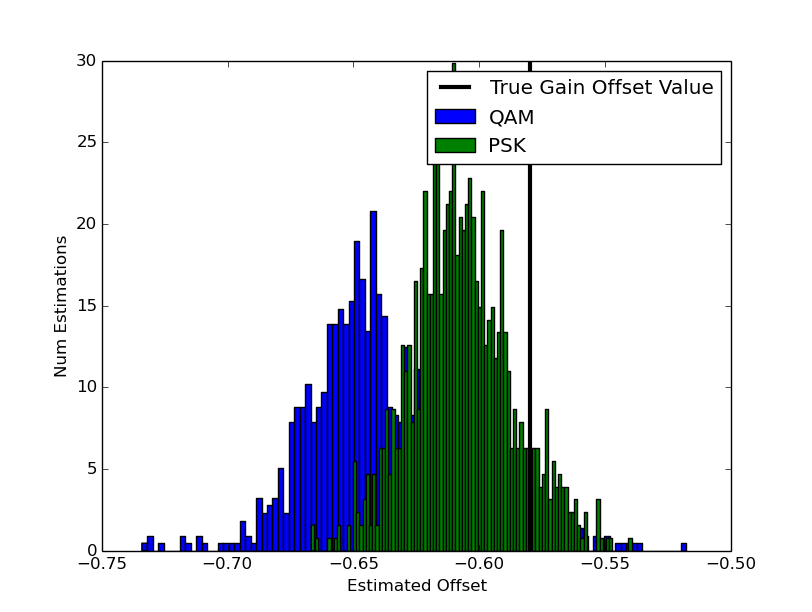}
	\caption{The histogram outputs for the PSK and QAM estimators both with true gain imbalance value of -0.58.}
	\label{fig:comparison}
    \vspace{-10pt}
\end{figure}

Figure \ref{fig:comparison} shows the importance of having separate decision boundaries for each modulation class. 
Though the true gain imbalance value of the input signal to the estimators is the same, the output histograms produced by the modulation-specific CNN gain imbalance estimators are not. 
This yields different decision boundaries for each modulation class.

As discussed in Section \ref{sec:est_results} and shown in Figures \ref{fig:gainSV_QAM} and \ref{fig:gainSV_PSK}, the CNN gain imbalance estimator trained for PSK showed to lower average sample variance than the CNN gain imbalance estimator trained for QAM. 
This resulted in fitted \emph{pdf}s with lower variance for PSK than QAM. 
Therefore, in general, lower minimum gain imbalances are needed to identify emitters with the same accuracy as the QAM estimators, as shown in Figures \ref{fig:snr_QAM} and \ref{fig:snr_PSK}, and more emitters can be identified uniquely.

\subsubsection*{Practical Considerations}
As expected, for both QAM and PSK, the lower the probability of mis-identification needed in a system, the higher the gain imbalance separation needed to achieve the needed level of accuracy. 
Therefore, in systems with a higher tolerance for mis-identification, more emitters can be uniquely identified, and in systems that require a low probability of mis-identification, fewer emitters can be uniquely identified. 
However, even in systems with a 20\% tolerance for mis-identification receiving signals exceeding $20dB$ SNR, emitters need to have a linear gain imbalance separation of at least 0.15. 
While few publications indicate measured gain imbalance values for real systems, most prior works in IQ imbalance estimation and compensation use test values on the order of 0.05 \cite{4389078, HsuSheen, 4102451, 5153348, 293640, 4753809, 713223}, indicating the gain imbalance values necessary to obtain even 80\% accuracy are not practical in real systems. 
Narrowing the range of gain imbalance values included in the training set would likely help combat this problem.   

The training set was simulated with gain imbalances between $[-0.9, 0.9]$, uniformly distributed, in order to incorporate any possible gain imbalance value the CNN estimator might encounter in a real system. 
However, training over such a large range has likely hindered the estimator's accuracy, as the network has had to learn to generalize over such a large range \cite{steve_milcom}. 
Given that $0.9$ is likely much larger than anything one might find in a real system, the training range could be narrowed to yield better results in estimator accuracy and therefore SEI ability.

\subsubsection*{Comparison to a Traditional Feature-Based Approach}
In \cite{Zhuo2017}, an SEI approach was developed using IQ imbalance estimates and Support Vector Machines (SVMs). 
The IQ imbalance estimation algorithm proposed uses statistical methods to determine gain and phase imbalance using the received symbols.
The approach also requires SNR estimation. 
They then plot gain versus phase imbalance in two dimensions and use SVMs to assign the received signal to an emitter. 

In order to provide an accurate comparison of the approach proposed in \cite{Zhuo2017}, five QPSK emitters were simulated with gain and phase imbalance values given in Table \ref{table:emitter_params}, for SNRs between [5dB, 35dB] at intervals of 5dB, and perfect synchronization has been assumed, as in \cite{Zhuo2017}. 
Additionally, the developed CNN gain imbalance estimator was retrained to accommodate these assumptions. 
More specifically, the CNN gain imbalance estimator was retrained using  training, validation, and test sets with SNRs between [0dB, 35dB], gain imbalance values $(\alpha)$ between $[0.0, 0.3]$, phase imbalance values $(\theta)$ between $[0^\circ, 5^\circ]$, and no frequency or sample rate offsets, in order to match the assumptions in \cite{Zhuo2017}.

\begin{table}[t]
\vspace{-10pt}
\centering
\caption{The simulated IQ Imbalance parameters used for comparison to the approach proposed in \cite{Zhuo2017}.}
\begin{tabular}{@{}cccccc@{}}
\toprule
 & Emitter 1 & Emitter 2 & Emitter 3 & Emitter 4 & Emitter 5 \\ \midrule
$\hspace{5pt} \alpha \hspace{5pt}$ & 0.1 & 0.13 & 0.15 & 0.17 & 0.19 \\
$\hspace{5pt} \theta \hspace{5pt}$ & $3^\circ$ & $3.3^\circ$ & $3.6^\circ$ & $3.9^\circ$ & $4.2^\circ$ \\ \bottomrule
\end{tabular}
\label{table:emitter_params}
\vspace{-10pt}
\end{table}

\begin{figure}[t]
\vspace{-20pt}
	\centering
	\includegraphics[width=1.0\linewidth]{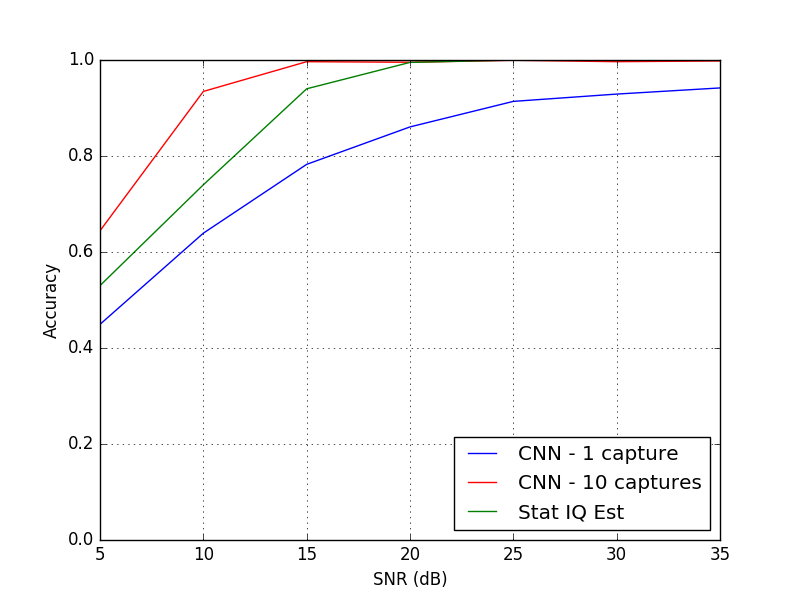}
	\caption{The accuracy of the developed SEI approach using CNN gain imbalance estimators compared to the accuracy of the approach proposed in \cite{Zhuo2017}, given one and ten captures of 1024 raw IQ samples.}
	\label{fig:method_comp}
\end{figure}

The results in Figure \ref{fig:method_comp} show the accuracy of the approach developed in this chapter to that proposed in \cite{Zhuo2017}, given one capture of 1024 raw IQ samples and given ten captures of 1024 raw IQ samples. 
When the approach developed in this chapter uses only one capture of 1024 raw IQ samples, the developed approach shows lower accuracies than the approach presented in \cite{Zhuo2017} by as much as 15\%.
However, it is important to note the developed approach uses only estimates of gain imbalance, while the approach presented in \cite{Zhuo2017} uses estimates of both gain and phase imbalance, providing more information about the emitter-of-interest. 
Phase imbalance estimates could also be incorporated into the approach developed in this chapter, and would likely increase performance. 
Furthermore, the approach presented in \cite{Zhuo2017} is dependent upon an estimate of SNR and assumes perfect synchronization, whereas the approach developed in this chapter needs no external measurements or estimates and can compensate for an imperfect receiver.

Additionally, the accuracy of the approach proposed in \cite{Zhuo2017} is calculated given 1330 symbols, whereas the accuracy of the approach developed in this chapter is given for one capture of 1024 raw IQ samples. 
Given multiple captures of raw IQ samples, the outputs can be aggregated, and the accuracy of the approach developed in this chapter improves, as shown in Figure \ref{fig:method_comp}. 
Therefore, given as few as ten captures of 1024 raw IQ samples, the accuracy of the approach developed in this chapter exceeds the performance of the approach proposed in \cite{Zhuo2017}, using less data and making far fewer assumptions.

\begin{figure}[t]
\vspace{-10pt}
	\centering
	\includegraphics[width=1.0\linewidth]{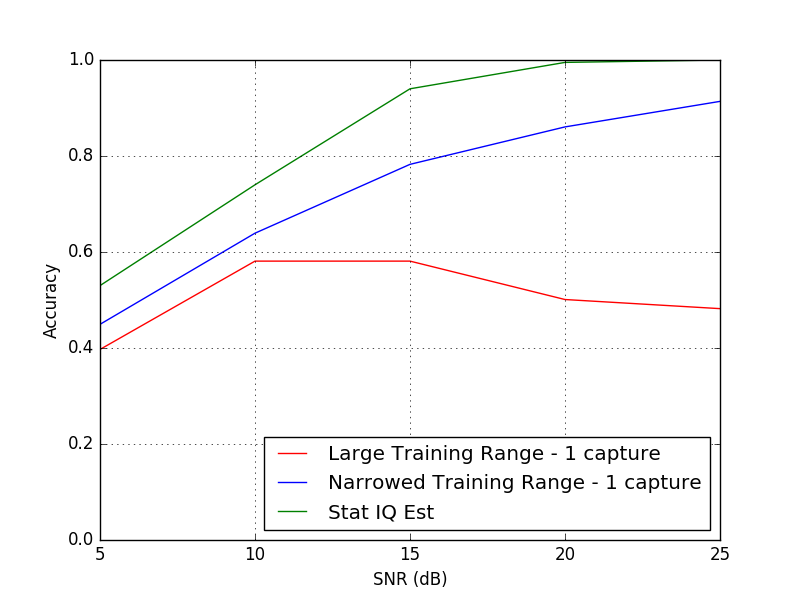}
	\caption{The accuracy of the developed SEI approach using the large training range described in Section \ref{sec:est_data} compared to the accuracy using the narrowed training range used to match the assumptions made in \cite{Zhuo2017}.}
	\label{fig:narrow_training}
    \vspace{-10pt}
\end{figure}

It should also be noted that limiting the range of IQ imbalance parameters and assuming perfect synchronization has improved the performance of the developed approach, as shown in Figure \ref{fig:narrow_training}. 
This confirms that training over smaller parameter ranges, more closely aligned with those one might find in a real system, would improve performance, and is consistent with the results and discussion in \cite{steve_milcom}. 

\section{Summary and Future Work}\label{sec:fw}
In this work, the capability of CNNs to estimate gain and phase imbalances between the in-phase and quadrature components of a signal was shown, assuming transmission through an AWGN channel. 
Performance analysis of the developed CNN IQ imbalance estimators, using QAM and PSK as test signals, showed the model to be modulation agnostic, and showed the model's ability to estimate both gain and phase imbalances, with performance increases as SNR and network input size increases. 
However, phase imbalance proved to be far more difficult to estimate than gain imbalance with the designed network architecture, showing much higher bias and sample variance values. 
Therefore, an SEI approach using parallel modulation-specific gain imbalance estimators was designed and evaluated. 

For both QAM and PSK modulation schemes, the proposed SEI approach showed increases in performance as SNR increases, in the form of smaller gain imbalance separations needed to achieve lower probabilities of mis-identification. 
Because the gain imbalance estimators trained for PSK slightly outperformed those trained for QAM, the proposed SEI approach performed better when the incoming signal was of a PSK modulation scheme. 
Though the approach was shown to need impractical gain imbalance separation values, even in high SNR scenarios, when the range of IQ imbalance parameters included in the training set is large, performance improved significantly when this range was narrowed.
Further, the accuracy of the approach was shown to exceed that of a traditional feature-based approach, given as few as ten captures of 1024 raw IQ samples. 

To improve the proposed SEI approach, the range of gain imbalance values included in the training set can be narrowed, so that the network has to generalize less. 
This was shown when comparing the developed approach to the approach in \cite{Zhuo2017}. 
The addition of more hardware impairments to the model, in order to further discriminate emitters, would also likely increase performance and is left for future work. 
Additionally, this work could be extended to further modulation schemes and to receiver IQ imbalance. 
Finally, the ability to estimate IQ imbalance and perform SEI in the presence of different channel models may be explored.

\bibliographystyle{IEEEtran}
\bibliography{bibliography}

\end{document}